\documentclass[12pt,a4paper,final]{iopart}
\usepackage{iopams}  
\usepackage{graphicx}
\usepackage[breaklinks=true,colorlinks=true,linkcolor=blue,urlcolor=blue,citecolor=blue]{hyperref}

 \newcommand{\bfE}{\mathbf{E}}

\newcommand{\bfB}{\mathbf{B}}

\newcommand{\bfJ}{\mathbf{J}}

\newcommand{\bfv}{\mathbf{v}}
\newcommand{\bfu}{\mathbf{u}}

\newcommand{\bfS}{\mathbf{S}} 
\newcommand{\bfQ}{\mathbf{Q}}
\newcommand{\tsP}{\mathbb{P}} 

\begin{document}

\title[Secondary reconnection in outflows]{Energy exchanges in reconnection outflows}

\author{Giovanni Lapenta}
\address{Center for Mathematical Plasma Astrophysics, Department of Mathematics, University of Leuven (KU Leuven), Celestijnenlaan 200B, 3001 Leuven, Belgium}
\ead{giovanni.lapenta@kuleuven.be}
\author{Martin V. Goldman, David L. Newman}
\address{University of Colorado, Boulder, CO 80309, USA}
\author{Stefano Markidis}
\address{High Performance Computing and Visualization (HPCViz), KTH Royal Institute of Technology, Stockholm, Sweden}

\begin{abstract}
Reconnection outflows are highly energetic directed flows that interact with the ambient plasma or with flows from other reconnection regions. Under these conditions the flow becomes highly unstable and chaotic, as any flow jets interacting with a medium. We report here massively parallel simulations of the two cases of interaction between outflow jets and between a single outflow with an ambient plasma. We find in both case the development of a chaotic magnetic field, subject to secondary reconnection events that further complicate the topology of the field lines. The focus of the present analysis is on the energy balance.  We compute each energy channel (electromagnetic, bulk, thermal, for each species) and find where the most energy is exchanged and in what form. The main finding is that the largest energy exchange is not at the reconnection site proper but in the regions where the outflowing jets are destabilizied.
\end{abstract}

\submitto{\PPCF}

\section{Introduction}
The reader will not be surprised to discover that reconnection is the process where magnetic energy is converted into kinetic energy, in the form of bulk   and thermal energy~\cite{biskamp}. The last decades have seen great advances in understanding the fundamental physics behind reconnection, a progress that has convinced NASA to launch the Multiscale MagnetoSpheric (MMS) mission \cite{mms-web}. With MMS, four closely spaced spacecraft can measure the local processes enabling reconnection with unprecedented spatial and temporal resolution \cite{Burchaaf2939}.

Alongside the progress made in understanding local reconnection physics, we need also to understand how reconnection acts on macroscopic scales to convert large fractions of energy. The current understanding of reconnection at the kinetic level is based on the formation of reconnection sites where two scales are present \cite{birn-priest}. In an outer region the ions become decoupled form the field lines and in an inner region the electrons become also decoupled allowing the topological break and reconnection of field lines.  The scale of the inner reconnection region is  the electron skin depth or gyro radius, depending on conditions~\cite{gembeta,hesse-guide}.

These electron scales are tiny in most systems of interest. In the Earth magnetosphere the electron scales are at the km scale, in the solar corona at the cm scale and in some laboratory plasmas at the micrometer to nanometer scale (depending on the type of experiment). How can large fractions of energy be converted if the region of reconnection are so tiny? Do we have to imagine all the energy needs to pass via these tiny portals? Obviously not. The past research has already indicated two ways out of this conundrum. 
 
 First, the energy conversion is not limited to the  region of field line breakage. Most of the energy is in fact processed as the plasma crosses the so-called separatrices~\cite{lapenta2014separatrices,marty-review}. These are the magnetic surfaces connected to the central reconnection site. In 2D models, this means that reconnection is topologically happening in a central point called x-point, in 3D this can extend to a so-called x-line or to more complex 3D topologies with null points \cite{priest-forbes}. 
 
 A well known implementation of this solution is  that of Petschek reconnection \cite{petschek}, based on the presence of anomalous resistivity in MHD models: energy is converted by  standing slow shocks at the separatrices. The process has an analogy in kinetic reconnection with strong energy conversion produced by the electric field caused by the Hall term at the separatrices \cite{shay2007two}. As the system size is enlarged to macroscopic scales, the kinetic process leads naturally to the formation of slow shocks and tangential discontinuities vindicating the validity of the Petschek model even within full kinetic treatment that does not rely on any ad hoc anomalous resistivity \cite{innocenti2015evidence}. 
 
 Second, there is the possibility of the presence of multiple reconnection sites. If each reconnection site is relatively small and can only process a small amount of energy, large energies can be processed if many reconnection sites are present. This can happen for example in turbulent systems where turbulence breaks up the reconnection process in a myriad reconnection sites \cite{matthaeus1986turbulent,1999ApJ...517..700L}. But also macroscopically laminar systems can have many reconnection sites via a process of  instability of a single reconnection site\cite{bulanov1979tearing,loureiro09,uzdensky10,skender}. A single reconnection site has a tendency of breaking up into many sites and progressively fill larger portions of the system \cite{lapenta2008self,bhattacharjee09}. The different reconnection site not only multiply the amount of energy processed but can also feed each other speeding up reconnection to increase even further the energy conversion rate \cite{lapenta2008self}.
 
 We present here a third possible explanation. The outflows of reconnection are highly energetic. The ions travel at Alfvenic speeds and the electrons at high superalfvenic speeds \cite{shay2011super,lapenta2013propagation}. As these flows interact with the ambient plasma or with outflows from neighboring reconnection sites, instabilities develop \cite{vapirev2013formation}. We have recently shown that these instabilities can create new reconnection sites where secondary reconnection processes start \cite{lapenta2015secondary}. We consider here the question of how much energy is converted locally and on average in the reconnection outflows when instabilities are observed.
 
 Of course the three mechanisms described above are not mutually exclusive. We can certainly assume that reconnection converts lots of energy by converting near the reconnection separatrices of multiple reconnection regions formed by the break up of an initial seed reconnection region and by the instabilities developing in the outflow from each reconnection site. The obvious tendency is for a spontaneous transition to turbulent reconnection \cite{lapenta-lazarian,bhattacharjee09}. A process we urgently need to investigate.
 
Below, section 2 shows examples of secondary reconnection sites in outflows from isolated reconnection sites interacting with pristine ambient plasmas. Section 3 shows the case of the interaction of outflows from neighboring reconnection sites. Section 4 describe the methodology applied in Sect. 5 to demonstrate that the outflow instabilities are powerful energy converters. 
Section 6 draws the conclusions of the present investigation.
 
 \section{Secondary reconnection from single reconnection region}
We consider first the case of a single isolated reconnection region. This case was considered in two previous publications. We reconsider here the same simulation parameters and the readers can find the detailed setup information in \cite{lapenta2014electromagnetic,lapenta2015secondary}. In  summary, we use the following parameters: $m_i/m_e=256$, $v_{th,e}/c=0.045$, $T_i/T_e=5$ . The initial state is a Harris layer of thickness $L/d_i=0.5$ with a density background $n_b=n_H/10$ with respect tot he peak Harris density and a guide field of $B_g=0.1B_0$. The domain has size $L_x=40d_i$, $L_y=15d_i$ and $L_z= 10d_i$. The initial gradients are along $y$ as is the reconnection inflow. The outflows of reconnection are along $x$ and $z$ is the initially ignorable coordinate where the 3D instabilities develop. Open boundary conditions in $x$ and $y$ allow free inflow and outflow, but $z$ is periodic. We avoid the question of reconnection onset (thereby justifying the use of a simple Harris sheet with background) by starting reconnection with a localized perturbation of the magnetic field that forms an x-line extending over the whole range of the $z$-axis. Around this x-line a reconnection region forms and  flow emerges jetting out along $x$. 
The simulations use the fully electromagnetic particle in cell code iPic3D~\cite{ipic3d} where both species are treated as particles.

These outflows with their embedded magnetic field move away from the central reconnection region and interact with the ambient plasma. The conditions of this interaction lead to instabilities. At large scales, these instabilities can be interpreted as interchange modes due to the presence of a stratified plasma surrounded by curved magnetic field lines \cite{nakamura2002interchange,guzdar2010simple,lapenta-bettarini-epl,lapenta2011self}. The curvature is unstable. But in the present case, the system is studied at smaller kinetic scales and the instability is more appropriately interpreted in terms of drift modes drawing their free energy from the drifts and density gradients present in the region of interaction between outflowing reconnected plasma and the ambient \cite{divin2015evolution,divin2015lower}.

\begin{figure}[htb!]
 \centering
\includegraphics[width=\columnwidth]{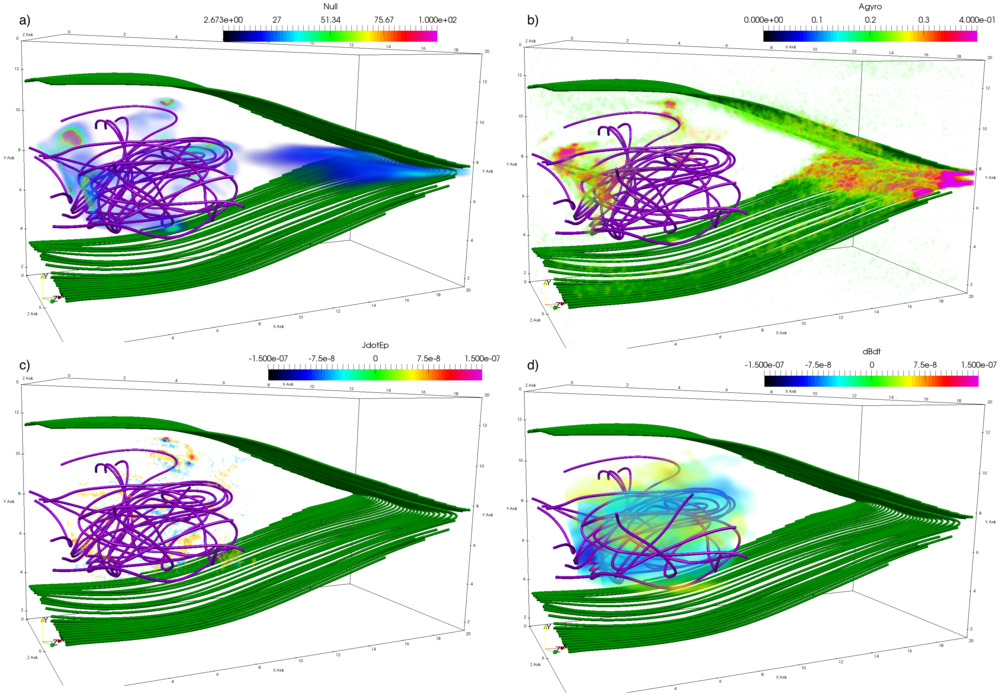}\\
 \caption{\label{figureone}False color volume rendering of the left half of the open system ($x<L_x/2$) with superimposed selected field lines emanating from the proximity of the central x-line. The top panels shows  the inverse of the magnetic field $B_0/B$ were $B_0$ is the lobe asymptotic field (a, left) and the agyrotropy (b, right). The bottom panels show  the dissipations in the frame moving with the center of mass velocity, $\bfJ \cdot (\bfE +\bfv_{CM} \times \bfB)$ where $\bfv_{CM}=(\bfv_i m_i + \bfv_e m_e)/(m_i+m_e)$ (c, left) and the rate of change of the magnetic energy, $\partial (B^2/2\mu_0)/\partial t$ (d, right). The field lines shown belong to two different sets. The green lines have just reconnected at the main central x-line (at $x=L_x/2$). The violet lines have a more twisted and chaotic topology acquired thanks to the instabilities in the reconnection outflows, and in particular due to secondary reconnection. The time shown is the end of the simulation,  $\omega_{ci}t=24.25$ (cycle 20000). }
\end{figure}

Figure \ref{figureone} shows the left half  of the system at the end of the simulation. To guide the eye we report a set of field lines emanating in proximity of $x=L_x/2$, just inside the separatrix surfaces. 

The region of interaction and development of the outflow instabilities is characterized by a chaotic magnetic field, with the presence of several regions of very small field (near nulls).  The inverse of the magnetic field (Fig. \ref{figureone}-a) is used to identify the regions of magnetic inversion. In kinetic simulations the fluctuations make the concept of a hard null fuzzy. We prefer to represent with volume rendering the regions where the magnetic field becomes very small, deep red saturating to pink being 1/100 of  the asymptotic field.
In a previous work~\cite{lapenta2015secondary}, these null regions were identified as regions of secondary reconnection where new magnetic connectivity can be shown among the field lines (shown here in violet). This conclusion is confirmed by the presence a strong agyrotropy in the components of the pressure tensor \cite{scudder2008illuminating}: the central x-line has  a strong agyrotropy as reported in previous 2D work~\cite{scudder2008illuminating} but the same levels of agyrotropy can be also found in several regions near the null of the magnetic field in the reconnection outflows.The processes around the null regions in the reconnection outflow can then be identified with secondary reconnection. 

These same regions are also the loci of the most intense energy exchange.  The two bottom panels of Fig. \ref{figureone} show the energy exchange between particles and fields and the rate of change of the magnetic energy. The dissipations are measured in the center of mass frame to exclude the energy expended simply in moving the plasma outward from the reconnection region: that is a macroscopic motion that does not involve any real dissipation. Once shifting to the center of mass frame~\cite{zenitani2011new,marty-review}, the true dissipation is illuminated. As can be observed the exchanges are bidirectional with magnetic energy being both created and destroyed (panel d) and work being done both by the particles and by the electric field (panel c). This bidirectional flow of energy has been reported also in satellite data observations taken in the outflow region of reconnection events\cite{hamrin2011energy}. While we are interested in overall net effects and in whether the generator and load regions (using the nomenclature in \cite{hamrin2011energy}) produce an net effect, the large energy exchange is important in its own right even if it were to even out: local observations can and have measured it. In Section IV, we will return to show that energy is indeed on average transferred from the magnetic field to the particles.

\section{Secondary reconnection within multiple reconnection regions}
As mentioned above, among the known mechanisms for expanding the rate of energy conversion by reconnection there is the plasmoid instability leading to the break up of a single reconnection region in several via the formation of plasmoids \cite{bulanov1979tearing,lapenta2008self,loureiro09}. When seen in 3D this means that flux ropes are formed in between reconnection points \cite{lapenta-bettarini-epl}.

\begin{figure}[htb!]
\includegraphics[width=\textwidth]{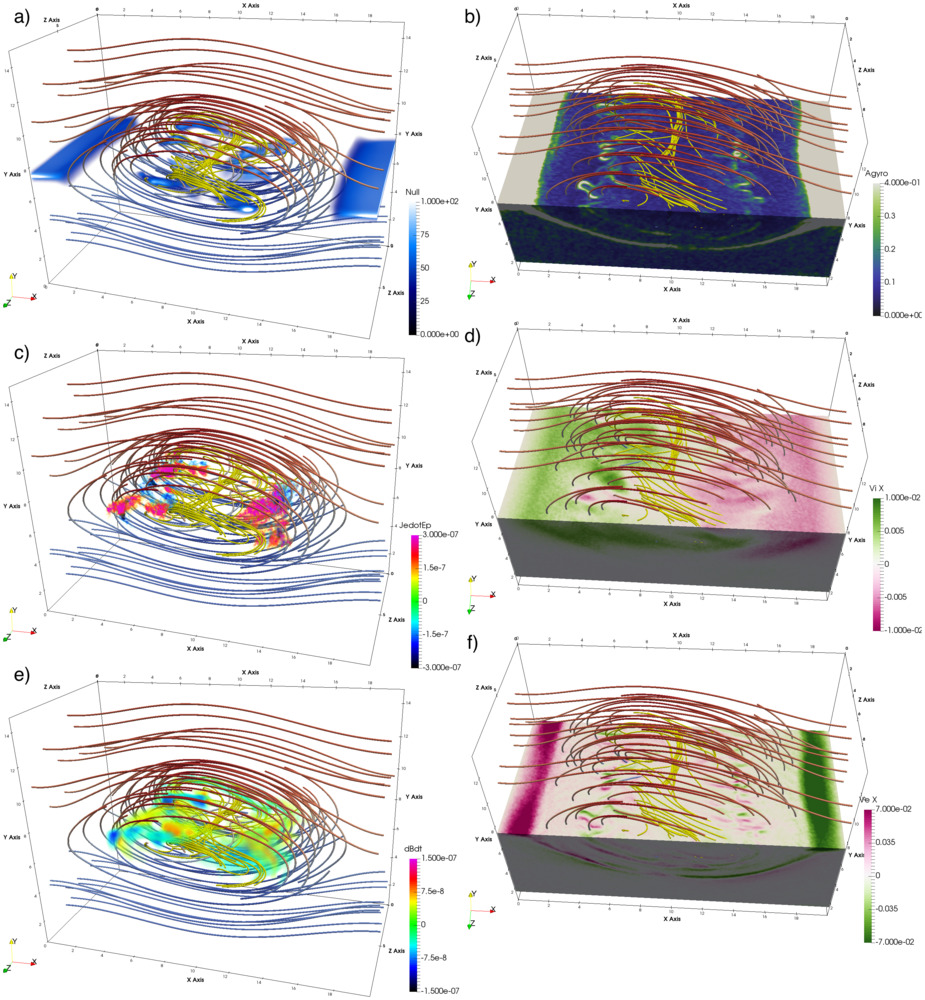}
 \caption{\label{rope1} False color volume rendering of the periodic system with superimposed selected field lines colored by the component $B_x$. The top panel shows the  inverse of the magnetic field $B_0/B$ were $B_0$ is the lobe asymptotic field (a, left) and the agyrotropy of the electron pressure tensor (b, right). The central panel shows the dissipations in the center of mass frame, $\bfJ \cdot (\bfE +\bfv_{CM} \times \bfB)$  (c, left) and the ion outflow speed (d, right). The bottom panels show the rate of magnetic energy change (e, left) and the electron outflow speed (f, right). The field lines shown belong to two different sets. The  lines colored by the component $B_x$ have just been reconnected once at the main  x-lines. The yellow lines have a more twisted and chaotic topology acquired thanks to the instabilities in the reconnection outflows, and in particular due to secondary reconnection.The time shown is the end of the simulation,  $\omega_{ci}t=18.19$ (cycle 15000).}
\end{figure}

This configuration can be replicated most economically in simulation by using periodic boundary conditions where one plasmoid is formed between two x-lines that in reality are the same repeated x-line in each period. Figure \ref{rope1} shows the results from this simulation. The parameters of the simulation are identical to those reported in \cite{lapenta2015secondary} for the flux rope case, except that the simulation is shifted by half a period to make the visualization of the flux rope more evident. In  summary, we use the same parameters and initial state as in the simulation above but the domain size has been reduced, only along the $x$ direction to  $L_x=20d_i$. The boundary conditions in $x$ are now periodic (while the others remain as above).  The initial perturbation causes two x-lines on the two edges of the box: they are really the same x-line via periodic boundary conditions.

Two flows emerge from each reconnection site flanking the central flux rope (Fig. \ref{rope1}-d). The two flows are bent by the modest 10\% guide field used in the simulation. The forming flux rope in this case is not only unstable to the same mechanisms observed for the single isolated reconnection site but it is also subject to the kink instability of the rope\cite{lapenta-bettarini-epl,markidis2014signatures} due to the presence of a larger azimuthal field compared with the axial (guide) field. These conditions are severely unstable according to the Kruskal-Shafranov condition and the flux rope kinks even in absence of drift effects from kinetic theory. However, in the present case the scales are again comparable to the ion inertial length and ion gyroradius so MHD theory cannot be used accurately and a full kinetic treatment is needed.

The instability in the center of the flux rope twists and bends the plasma and magnetic field leading to a multiplicity of null regions where secondary reconnection again develops.  As in the case of open boundary simulation with a single isolated x-line, a detailed analysis uncovers that where the magnetic field reaches a null secondary reconnection develops\cite{lapenta2015secondary}.  Figure \ref{rope1} reports the state of the simulation at the end of the run. The regions of null field (panel a) show also a strong agyrotropy (panel b). Other indicators of reconnection  concur in proving that the yellow field lines have been subject to a topological restructuring by secondary reconnection \cite{lapenta2015secondary}. At the time shown, these processes have led to the formation of a twisted flux rope (yellow field lines). Again these regions show powerful energy exchanges when the dissipations (panel c) and the rate of magnetic energy change (panel e) are considered.

Recently,  the first possible experimental confirmation of this process has been published \cite{oieroset2016mms}. The MMS fleet of four spacecraft has observed a situation where reconnection happens between two converging jets originated by two neighbouring reconnection sites. The conditions suggested by the observations are identical to those observed in Fig.~\ref{rope1}-c  This is exactly the case predicted in \cite{lapenta2015secondary}  and reconsidered here. The MMS observation can be interpreted just as the results  in Figure \ref{rope1}-c where the ion flows from the two neighbouring reconnection x-line are observed to interact. In the region of their interaction, secondary reconnection produces new reconnection outflows on smaller scales, visible in the electron flow speed in Fig.~\ref{rope1}-f. 

\section{Methodology for the analysis of the energy balance}
To assess the  overall energy output from reconnection we consider an integral along the $z$ and $y$ directions and observe the local energy deposition at different locations away form the central x-line positioned at $x=L_x/2$. This integral considers all energies exchanged in each $x={\rm const}$ plane:
\begin{equation}
\overline{\Psi}(x) = \frac{1}{L_z} \int_0^{L_z} \int_0^{L_y} \Psi(x,y,z) dy dz
\end{equation}
where $\Psi$ is the generic energy contribution. As shown above some contributions alternate in sign and the integral measure considers the net contribution once all fluctuations are eliminated by the integral. 
The energy deposition at different $x$ gives information on where the most net energy is deposited. 

Another concept of interest is the overall energy deposited within the volume around the reconnection region. To investigate this overall energy budget, we consider a cumulative integral over $x$ starting from the $x$ point and integrating in either directions along x: 
\begin{equation}
\overrightarrow{\Psi}(x) = \frac{1}{L_z}  \int_0^{L_z} \int_0^{L_y} \int_{x-point}^x\Psi(x^\prime,y,z) dx^\prime dy dz
\end{equation}
computed rightward of the x-point and analogously:
\begin{equation}
\overleftarrow{\Psi}(x) = \frac{1}{L_z}  \int_0^{L_z} \int_0^{L_y} \int_{x}^{x-point}\Psi(x^\prime,y,z) dx^\prime dy dz
\end{equation}
computed leftward of the x-point (at $x=L_x/2$).

We consider each energy balance equation:  electromagnetic, bulk and thermal energy for each species.  The electromagnetic energy is given by the energy of the magnetic and electric field, the first typically dominating in space reconnection processes. The bulk energy is the energy associated with the bulk velocity of a species.
Obviously, in kinetic models there is no specific fluid but still we can define the local average bulk speed averaged over all particles of a species in the proximity of a point (e.g. within a cell), $\bfu$. The bulk energy is defined as
$U_{bulk}=nm_pu^2/2$ with particle mass $m$, particle density $n$ and bulk
speed $u$ of a species $s$ (the species index is dropped). 
The thermal energy is then given by the energy of the velocity fluctuations of each individual particle with respect to the local average bulk speed. The thermal energy is defined as
$U_{th}=Tr \tsP/2$ based on the trace of the pressure tensor. 

  The balance of  electromagnetic
energy in any system is given by:

 \begin{equation}
{\frac{1}{2}\frac{\partial }{\partial t}\left( \epsilon_0 E^2
+\frac{1}{\mu_0}B^2 \right) = -\bfE \cdot \bfJ -  \nabla \cdot \bfS}
\label{elm-energy-balance}
 \end{equation} 
 where $\bfS=\bfE \times \bfB/\mu_0$ is the Poynting vector.

For the species, it is interesting to consider two different forms of the balance equations: in the  Eulerian frame and in the Lagrangian frame moving with the bulk speed of each species. The Eulerian frame considers the balance as it would be seen by a local observer at rest in the reference frame. This is for example the case of a spacecraft  (the speed of the spacecraft is typically negligible compared with the observations time and the speed of motion of the  species). 
The Lagrangian frame moving with the species observes the energy exchanges as they would be seen by an element of the "fluid" moving at the speed of the species considered. 

Starting with the Eulerian frame, the balance of  bulk energy in a system is given by~\cite{burgers1969flow}
\begin{equation} 
{\frac{\partial U_{bulk}}{\partial t} = \bfE \cdot \bfJ
-  \nabla \cdot \bfQ_{bulk} - \bfu \cdot \nabla \cdot \tsP}
\label{bulk-energy-balance}
 \end{equation} 
 where $\bfQ_{bulk}=\bfu U_{bulk}$ is
the bulk energy flux and $\tsP$ is the pressure tensor.
The
balance of  thermal energy is given by 
\begin{equation}
{\frac{\partial U_{th}}{\partial t} = -  \nabla \cdot \bfQ_{th} - 
\nabla \cdot \bfQ_{hf} + \bfu \cdot \nabla \cdot \tsP}
\label{th-energy-balance} 
\end{equation} where 
$
\bfQ_{th}= \bfu \cdot \tsP 
$
is
the internal energy flux and $\bfQ_{hf}$ is the heat flux. The two energy equations can be summed to express the total energy change in terms of the enthalpy flux  $\bfQ_{enth}=\bfu U_{th} + \bfu \cdot \tsP $

For the Lagrangian frame, two different frames are used for ions and electrons. The two Eulerian equations above can be rewritten using 
the advection derivative 
${D}/{Dt}={\partial}/{\partial t} +\bfu \cdot \nabla$
with the velocity for each species used for defining the frame.

The Lagrangian Bulk Energy Equation is:
\begin{equation} 
{\frac{D U_{bulk}}{D t} = \bfE \cdot \bfJ -U_{bulk} \nabla \cdot \bfu - \bfu \cdot \nabla \cdot \tsP}
\label{bulk-energy-balance-lagrange-final}
 \end{equation} 
 and the Thermal Energy Equation is: 
\begin{equation}
{\frac{D U_{th}}{D t} = -U_{th} \nabla \cdot \bfu - \tsP : \nabla \bfu}
\label{th-energy-balance-lagrangian-final} 
\end{equation} 

\section{Results for the energy balance in the outflow}
Each term of the equations above is analyzed for the run above with open boundary conditions for a single isolated reconnection region. 
In all figures below each term of the right hand side of the Eulerian equations are plotted. 
 
\begin{figure}[htb!]
\begin{tabular}{cc}
  \includegraphics[width=.5\columnwidth]{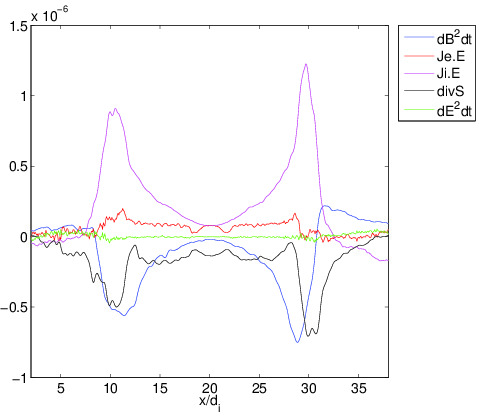}&
  \includegraphics[width=.5\columnwidth]{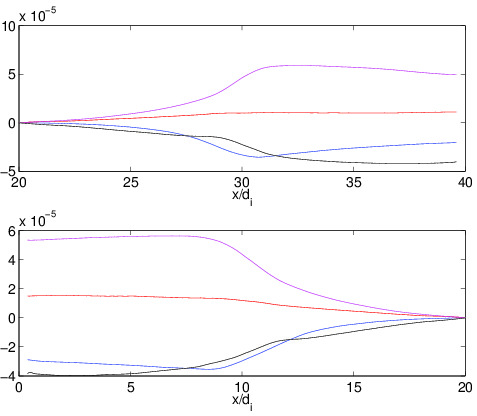}
\end{tabular}
\caption{Electromagnetic energy balance: Top: averaged over z and integrated over y. Middle and bottom: cumulative integral along x rightward (middle) and leftward (bottom). The two field energies include the correct normalization factors, not indicated by the labels. The time shown is cycle $\omega_{ci}t=21.82$ (cycle 18000).}
\label{elm_balance}
\end{figure}

The electromagnetic energy balance is reported in  Fig. \ref{elm_balance}. The dominant work is done by the ions and it is done predominantly in the fronts. For the electrons the work is more uniformly spread. Neither are especially concentrated near the central x-line, in fact the energy exchange there is a minute fraction.  The Poynting flux term and the actual conversion of magnetic energy nearly equipartition the energy: the energy spent by the electric field on the particles comes in comparable fraction from the divergence of the Poynting flux (i.e. electromagnetic energy arriving form other locations) and from the local decrease in magnetic energy. The change in electric field energy is negligibly small.  

A remarkable left-right asymmetry develops. The stochastic nature of the initial noise in PIC simulations amplified by the growth of the instabilities on the two fronts leads to a chaotic behaviour. We have tested that if the same simulation is repeated with a different compiler on a different hardware (running the same exact case at NASA Pleiades and CEA Curie) leads to qualitatively similar beheviour with completely different detailed features. This is the nature of chaos in non-linear system. Another proof that the magnetic field in the reconnection outflow is chaotic.

\begin{figure}[htb!]
\noindent Electrons:\\
\begin{tabular}{cc}
  \includegraphics[width=.5\columnwidth]{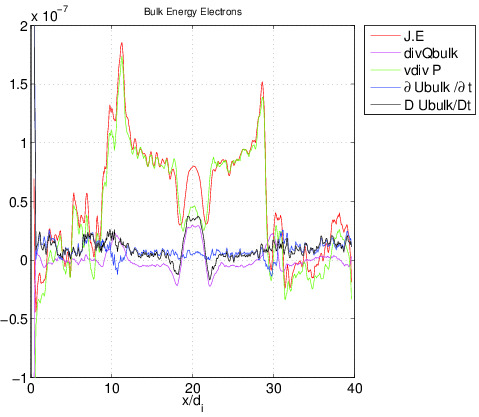}&
  \includegraphics[width=.5\columnwidth]{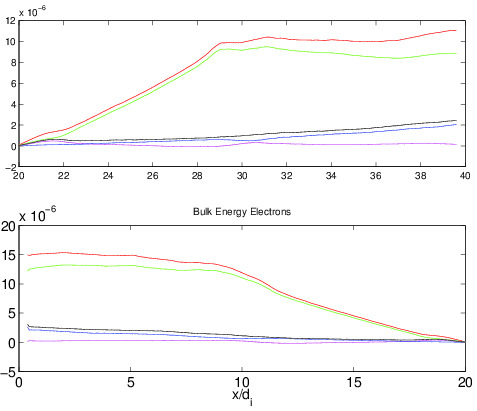}
\end{tabular}\\
 \noindent Ions:\\
\begin{tabular}{cc}
  \includegraphics[width=.5\columnwidth]{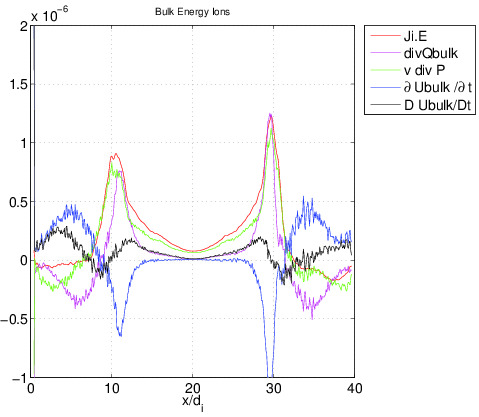}&
  \includegraphics[width=.5\columnwidth]{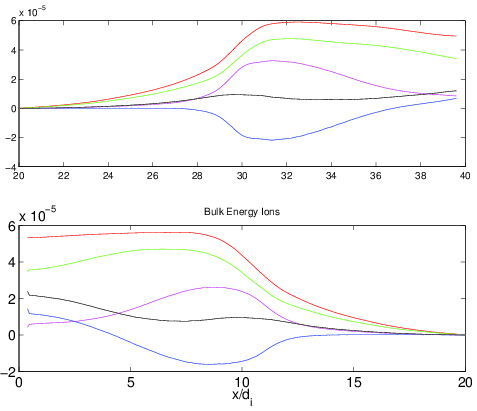}
\end{tabular}
\caption{Electron and Ion bulk energy balance at $\omega_{ci}t=21.82$: Left: averaged over z and integrated over y.  
Right: cumulative integral along x rightward (upper) and leftward (lower). The top panels are the electrons and the lower panels the ions. The time shown is cycle 18000.}
\label{bulk_balance}
\end{figure}

The analysis of the bulk energy is in Fig.  \ref{bulk_balance}. The work done by the fields on the particles is predominantly going to   $\bfu \cdot \nabla \cdot \tsP$. This is  the macroscopic work that has pushed generations of travelers sitting on trains pulled by steam locomotives. The fields are moving the plasma outward towards the ambient plasma. The divergence of the bulk energy flow shows two remarkable features: the central x-line (only for the electrons) and the two fronts.  In these regions the particles gain most of their bulk energy.  The ions gain the lion's share: the electrons and ion at the front move at the same speed, the ions inertia being obviously larger.

In the Lagrangian frame the ions are constantly accelerated (i.e. the Lagrangian derivative of their energy increases) towards the fronts where their acceleration stops.   The electrons have a similar behaviour but with a stronger acceleration at the x-line, nearly absent for the ions. In the Eulerian frame the inertia is not part of the temporal derivative and the inertial effects are missed. For a near steady state, the Eulerian derivative should be zero but here we have a moving front. In the vicinity of the central x-line where a near steady state is present, the Eulerian derivative is nearly zero. At the fronts goes from negative (where energy is moving away) to positive (where energy is moving to). This is merely a motion of the front, the Lagrangian frame should be used for a more insightful interpretation.

\begin{figure}[htb!]
\noindent Electrons:\\
\begin{tabular}{cc}
  \includegraphics[width=.5\columnwidth]{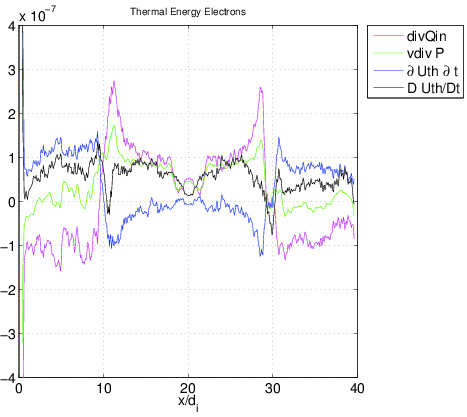}&
  \includegraphics[width=.5\columnwidth]{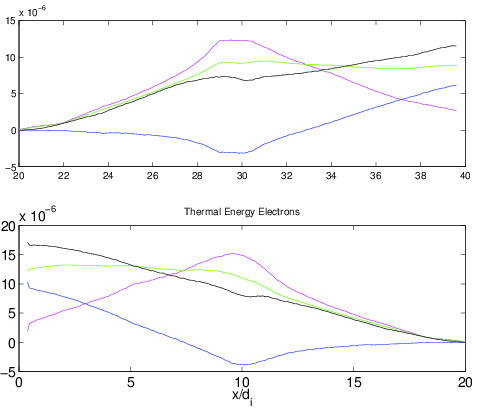}
  \end{tabular}\\
 \noindent Ions:\\
\begin{tabular}{cc}
  \includegraphics[width=.5\columnwidth]{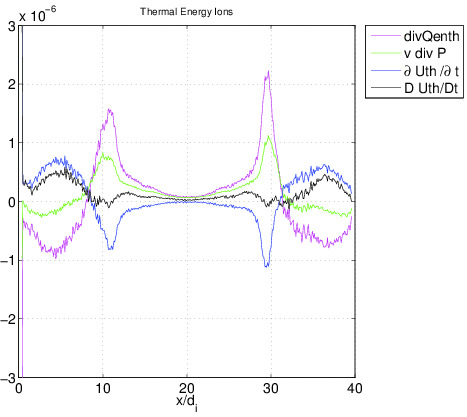}&
  \includegraphics[width=.5\columnwidth]{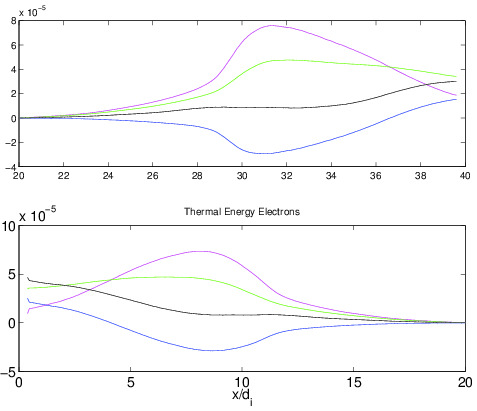}
\end{tabular}
\caption{Electron and ion thermal energy balance at $\omega_{ci}t=21.82$. Left: averaged over z and integrated over y.  
Right: cumulative integral along x rightward (upper) and leftward (lower). The top panels are the electrons and the lower panels the ions.}
\label{thermal_balance}
\end{figure}

The thermal energy (Fig. \ref{thermal_balance}) is dominated by the derivative of the internal energy flux. That term is partly balanced by the $\bfu \cdot \nabla \cdot \tsP$ term that couples to the bulk energy balance (the energy used by the locomotive to push the train is taken form the enthalpy of the locomotive's steam). 

As in the case of the bulk energy, the Lagrangian derivative of the thermal energy is readily interpreted as a constant heating of the species, dropping at the front. The Eulerian derivative is again less transparent, for the same reasons outlined above. 

An interesting point in all the energies above is that there is significant energy exchange also downstream of the fronts. There is in fact interaction between outer particles reflected by the moving fronts with the ambient plasma  giving an impact even before the front actually arrives. This mechanism is present for both species.
\section{Conclusions}

The process of reconnection produces a highly energetic jet. When reconnection happens in a isolated site the jet interacts with the surrounding plasma, when multiple reconnection sites are in proximity, jets from neighbouring sites interact. In both situations, a jet becomes destabilized leading to a chaotic magnetic field topology in the region of interaction. Within it, secondary reconnection develops and strong energy exchange takes place. The local rate of change of the magnetic energy and the work between the particles and the electric field is stronger by one order of magnitude in the region of the interaction of the of the outflows than in the reconnection region proper.
 
 The energy exchanges in the outflows are very strong but are also fluctuating with energy going both from the particles to the field (dynamo or generator) or from the field to the particle (reconnection or load). 
 
 To assess the actual energy budget, we analyzed the integrated energy balance at each distance away from the reconnection site along the direction of the outflow (direction $x$), integrated in the other directions. Two sets of equations are written: in a Lagrangian and in a Eulerian frame. The Lagrangian frame is of more direct physical relevance as it determines how the plasma species are actually exchanging energy. The Eulerian frame is convenient for comparison with in situ local measurements.

The main conclusion is that the front is the region of most intense energy exchange. All the main indicators of net energy transfer reach their peak there: work done by the electromagnetic fields, rate of magnetic energy decrease and divergence of the Poynting flux, divergence of bulk energy and thermal energy (together forming the enthalpy flux)  of each species.

\section*{Acknowledgments}
The present work is supported by  the NASA MMS Grant No. NNX08AO84G, by the Onderzoekfonds KU Leuven (Research Fund KU Leuven),
by the Interuniversity Attraction Poles Programme of the Belgian Science Policy
Office (IAP P7/08 CHARM)  and by the DEEP-ER project of the European Commission. The simulations were conducted on  NASA (NAS and NCCS) supercomputers, on  DOE-NERSC supercomputers, at the  VSC Flemish supercomputing centre and on the  facilities provided by  PRACE research infrastructure Tier-0 grants.

\section*{References}
\bibliographystyle{iopart-num}
\bibliography{bibliografie} 
\end{document}